\documentclass[a4paper,12pt]{article}
\usepackage{epsfig}
\usepackage{textcomp}
\usepackage{sparticles}     
\usepackage{latexsym}

\begin{document}

\newcommand{\Pom}{\mbox{{\it I$\!$P}}}

\parskip=0.3cm

\vskip 0.5cm \centerline{\bf A Deeply Virtual Compton Scattering
Amplitude} \vskip 0.3cm \centerline{M.~Capua$^{a\diamond}$,
S.~Fazio$^{a\ast}$, R.~Fiore$^{a\ddagger}$, L.~Jenkovszky
$^{b\dagger}$, and F.~Paccanoni$^{c\star}$}

\vskip 1cm

\centerline{ $^a$ \sl Dipartimento di Fisica, Universit\'a della
Calabria} \centerline{ \sl Istituto Nazionale di Fisica Nucleare,
Gruppo collegato di Cosenza} \centerline{  \sl I-87036 Arcavacata
di Rende, Cosenza, Italy} \centerline{$^b$ \sl Bogolubov Institute
for Theoretical Physics, National Academy of Sciences of Ukraine}
\centerline{\sl UA-03143 Kiev, Ukraine} \centerline{$^{c}$ \sl
Dipartimento di Fisica, Universit\'a di Padova} \centerline{ \sl
Istituto Nazionale di Fisica Nucleare, Sezione di Padova}
\centerline{ \sl via F. Marzolo 8, I-35131 Padova, Italy}

\vskip 0.1cm

\begin{abstract}
A factorized Regge-pole model for deeply virtual Compton
scattering is suggested. The use of an effective
logarithmic Regge-Pomeron trajectory provides for the description
of both ``soft'' (small $|t|$) and ``hard'' (large $|t|$) dynamics.
The model contains explicitly the photoproduction and the DIS
limits and fits the existing HERA data on deeply virtual Compton scattering.
\end{abstract}

\vskip 0.1cm

$
\begin{array}{ll}
^{\diamond}\mbox{{\it e-mail address:}} &
   \mbox{capua@cs.infn.it} \\
^{\ast}\mbox{{\it e-mail address:}} &
   \mbox{fazio@cs.infn.it} \\
^{\ddagger}\mbox{{\it e-mail address:}} &
   \mbox{fiore@cs.infn.it} \\
^{\dagger}\mbox{{\it e-mail address:}} &
   \mbox{jenk@bitp.kiev.ua} \\
^{\star}\mbox{{\it e-mail address:}} & \mbox{paccanoni@pd.infn.it}
\end{array}
$


\section{Introduction} \label{s1}

Interest in deeply virtual Compton scattering (DVCS)
$ep\rightarrow e\gamma p$ is related to the prospects to use
it as a tool in studies of
Generalized Parton Distributions (GPD)~\cite{Mueller,DVCS}.

\begin{figure}[h]
\begin{center}
\includegraphics[clip,scale=1.0]{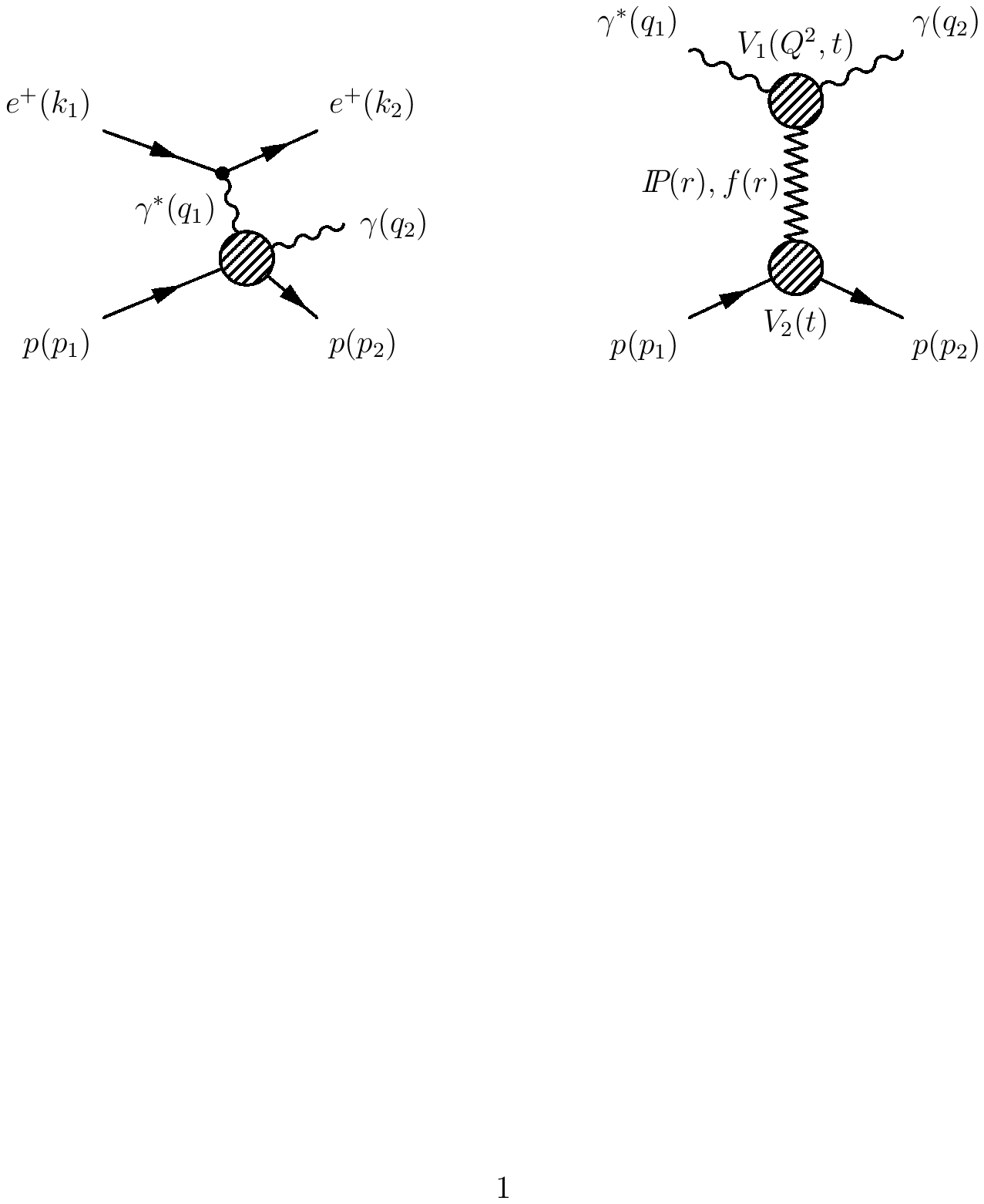}
\end{center}
\vspace{-10cm}
\caption{\small\it{
a) Diagram of a DVCS event at HERA; b) DVCS
amplitude in a Regge-factorized form.
}}
\label{fig:diagram2}
\end{figure}

At HERA the DVCS cross-section has been
measured~\cite{H1_05,ZEUS}, in diffractive $ep$
interactions, as a function of $Q^2$, $W$ and $t$ that are
respectively the photon virtuality, the invariant mass of the
$\gamma ^*p$ system and the squared 4-momentum transferred at the
proton vertex; the diagram in Fig.~\ref{fig:diagram2}a shows the
production of a real photon at HERA.

The $Q^2$ evolution of the DVCS amplitude has been
studied in several papers, mainly in the context of perturbative
quantum chromodynamics (QCD)~\cite{DD, QCD} and recently in~\cite{Muellerfit}. 
The $t$ dependence in many papers
was introduced by a simple factorized exponential in
$t$, which however differs from the Regge pole theory. Since the
electron-proton scattering at HERA is dominated by a single photon
exchange, the calculation of the DVCS scattering amplitude reduces
to that of the $\gamma^*p\rightarrow\gamma p$ amplitude, 
which at high energies, in the
Regge pole approach, is dominated by the exchange of
positive-signature Reggeons, associated with the Pomeron and the
$f$-trajectories~\cite{Collins}. This DVCS amplitude is shown in
Fig.~\ref{fig:diagram2}b in a Regge-factorized form. In the figure
$q_{1,2}$ are the four-momenta of the incoming and outgoing
photons, $p_{1,2}$ are the four-momenta of the incoming and
outgoing protons, $r$ is the four-momentum of the Reggeon
exchanged in the $t$ channel, $r^2=t=(q_1-q_2)^2$ and
$s=W^2=(q_1+p_1)^2$ is the squared centre-of-mass energy of the incoming
system.

\vspace{1cm}

Unless specified (as in the deep inelastic scattering (DIS) limit, 
discussed in Sec.~3),
$q_2^2=0$, and hence, for brevity, $q^2_1=-Q^2$. In the upper
vertex $V_1$, Fig. 1b, a virtual photon with 4-momentum $q_1$, and
a Reggeon (e.g. Pomeron) with 4-momentum $r$, enter and a real
photon, with 4-momentum $q_2=q_1+r$ appears in the final state as
an outgoing particle. The vertex $V_1$ depends on all the possible
invariants constructed with the above 4-momenta,
$V_1[q_1^2,r^2,q_1\cdot r],$ where $r^2=t\leq 0, q_1^2=-Q^2\leq
0.$ The three invariants are not independent since the mass-shell
condition for the outgoing photon, $q_2^2=(q_1+r)^2=0$, provides
the relation
\begin{equation}\label{zeta01}
q_1\cdot r={-q_1^2-r^2\over 2}={Q^2-t\over 2}.
\end{equation}

Hence, the vertex can be considered as a function of the
invariants $[Q^2,q_1\cdot r]$ or $[t,q_1\cdot r]$. This does not
mean that the variables cannot appear separately  but it could
also happen that $q_1\cdot r$ become a scaling variable, and
consequently the vertex will finally depend on $q_1\cdot r$ only. 
It depends on the dynamics of the process and,
for the moment, we prefer to keep $t,$ apart from $Q^2,$ as the
second independent variable.

Electroproduction of a vector meson gives another example since in
this case $(q_1+r)^2=M_V^2$, and the variable $q_1\cdot r$ becomes
\begin{equation}\label{zeta02}
q_1\cdot r={M_V^2-q_1^2-r^2\over 2}={M_V^2+Q^2-t\over 2}.
\end{equation}

The interplay of the $Q^2$- and $t$-dependence in the DVCS
amplitude was recently discussed in Ref.~\cite{FJMP1}, where the
existence of a new, universal variable $z$ was suggested. The
basic idea is that $Q^2$ and $t$, both having the meaning of a
squared mass of a virtual particle (photon or Reggeon), should be
treated on the same footing, by means a new variable, defined as
\begin{equation}\label{zeta}
z=q_1^2+t=-Q^2+t,
\end{equation}
 in the same way as the vector meson
mass squared is added to the squared photon virtuality, giving
$\tilde{Q^2}=Q^2+M_{V}^{2}$ in the case of vector meson
electroproduction~\cite{INS, MPP}.

In this paper we examine an explicit model for DVCS with $Q^2$-
and $t$-dependences determined by the $\gamma^* \Pom \gamma$
vertex. We suggest the use of the new variable defined in Eq.~\ref{zeta}
with its possible generalization to vector meson electroproduction,

\begin{equation}\label{zeta2}
z=t-(Q^2+M^2_V)=t-\tilde{Q^2}
\end{equation}
or virtual photon (lepton pair) electroproduction,
\begin{equation}\label{zeta2}
z=t-(Q_1^2+Q_2^2),
\end{equation}

\noindent
where $Q_2^2=-q_2^2$.
However, differently from Ref.~\cite{FJMP1}, here we
introduce the new variable only in the upper, $\gamma^* \Pom
\gamma$ vertex, to which the photons couple. 

In the next Section we introduce the model. Its viability is
supported by the correct photoproduction- ($Q^2=0$) and DIS-
($Q^2>0$ and $t\rightarrow 0$) limits, demonstrated in Sec. 3.
Fits to the data are presented in Sec.~4, while discussions and
conclusions are in Sec.~5.

\section{The model} \label{s2}

According to Fig.~\ref{fig:diagram2}b, the DVCS amplitude can be written as
\begin{equation}\label{A1}
A(s,t,Q^2)_{\gamma^* p\rightarrow\gamma
p}=-A_0V_1(t,Q^2)V_2(t)(-is/s_0)^{\alpha(t)},
\end{equation}
where $A_0$ is a normalization factor, $V_1(t,Q^2)$ is the $\gamma^*
\Pom \gamma$ vertex, $V_2(t)$ is the $p \Pom p$ vertex and $\alpha(t)$ is the
exchanged Pomeron trajectory, which we assume in a logarithmic form:
\begin{equation}\label{alpha}
\alpha(t)=\alpha(0)-\alpha_1\ln(1-\alpha_2 t).
\end{equation}

Such a trajectory is nearly linear for small $|t|$, thus
reproducing the forward cone of the differential cross section,
while its logarithmic asymptotics provides for the large-angle
scaling behavior~\cite{CGTB, FJMP2}, typical of hard collisions at
small distances, with power-law fall-off in $|t|$, obeying quark counting
rules~\cite{CGTB, MMT, Farrar}. Here we are referring to the
dominant Pomeron contribution plus a secondary trajectory, e.g.
the $f$-Reggeon. Although we are aware of the importance of this
subleading contribution at HERA energies, nevertheless we cannot
afford the duplication of the number of free parameters, therefore
we include it effectively by rescaling the parameters. Ultimately,
the Pomeron and the $f$-Reggeon have the same functional form,
differing only by the values of their parameters. Furthermore, the
Pomeron~\cite{BFKL} itself is unlikely to be a single term, so
instead of including several Regge terms with many free
parameters, it may be reasonable to comprise  them in a single
term, called \textgravedbl effective Reggeon\textacutedbl or
\textgravedbl effective Pomeron\textacutedbl, depending on the
kinematical region of interest. Although the parameters of this
effective Reggeon (Pomeron) (e.g. its intercept and slope) can be
close to the \textgravedbl true\textacutedbl one (whose form is at
best a convention), for the above reason they never should be
taken as granted.

For convenience, and following the arguments based on duality (see
Ref.~\cite{BGJPP} and references therein), the $t$ dependence of
the $p \Pom p$ vertex is introduced via the $\alpha(t)$ trajectory:
$V_2(t)=e^{b \alpha(t)}$ where $b$ is a parameter. 
A generalization of this concept will be
applied also to the upper, $\gamma^* \Pom \gamma$ vertex by
introducing the \textgravedbl trajectory\textacutedbl
\begin{equation}\label{beta}
\beta(z)=\alpha(0)-\alpha_1\ln(1-\alpha_2 z),
\end{equation}
where the value of the parameter $\alpha_2$ may be different in
$\alpha(t)$ and $\beta(z)$ (a relevant check will be possible when more
data will be available). Hence the scattering amplitude (6), with the
correct signature, becomes

\begin{equation}\label{A2}
A(s,t,Q^2)_{\gamma^* p\rightarrow\gamma p}= -A_0e^{b\alpha(t)}e^{b
\beta(z)}(-is/s_0)^{\alpha(t)}= -A_0e^{(b+L)\alpha(t)+b\beta(z)},
\end{equation}
where $L\equiv\ln(-is/s_0)$.

The model contains a limited number of free parameters. Moreover,
most of them can be estimated a priori. The product
$\alpha_1\alpha_2$ is just the forward slope $\alpha'$ of the
Reggeon ($\approx 0.2$~GeV$^{-2}$ for the Pomeron, but much higher
for $f$ and/or for an effective Reggeon) \footnote{As emphasized
in a number of papers, e.g. in Ref.~\cite{FJPP}, the wide-spread
prejudice of the \textgravedbl flatness\textacutedbl of the
Pomeron in electroproduction is wrong for at least two reasons:
one is that it was deduced by fitting data to a particular
\textgravedbl effective Reggeon\textacutedbl (see the relevant
discussion above) and the second is that the Pomeron is universal,
and its nonzero slope is well known from hadronic reactions.}. The
value of $\alpha_1$ can be estimated from the large-angle quark
counting rules~\cite{CGTB, MMT, Farrar}. For large $t$ ($|t|>>$1~GeV$^2$)
the amplitude goes roughly (a detailed treatment of this
point can be found in Refs.~\cite{CGTB, FJMP2}) like $\sim
e^{-\alpha_1\ln(-t)}=(-t)^{\alpha_1},$ where the power $\alpha_1$
is related to the number of quarks in a collision~\cite{CGTB, MMT,
Farrar}, e.g. their number minus one. Various versions of the
counting rules suggest different combinatorics giving slightly
different values for this power. We set $\alpha_1=1,$ and
hence $\alpha_2=\alpha'$. For the intercept of our
effective Reggeon, dominated by the Pomeron, we set
$\alpha(0)=1.25$ as an \textgravedbl average\textacutedbl~over the
\textgravedbl soft+hard\textacutedbl~Pomerons \footnote {This is
an obvious simplification and we are fully aware of the variety of
alternatives for the energy dependences, e.g. that of a dipole
Pomeron, as in Ref.~\cite{BGJPP}, a \textgravedbl soft\textacutedbl plus
a \textgravedbl hard\textacutedbl one, as e.g. in Ref.~\cite{DD}.
Ultimately, from QCD's BFKL equation~\cite{BFKL} an infinite
number of Pomeron singularities follows unless simplifications are
used. For the present study in term of the new, $z$, variable the simplest
\textgravedbl supercritical\textacutedbl~Pomeron~\cite{DD} with an
effective intercept is suitable.}.
The above values of the parameters should not be taken as granted,
they should be considered as starting values in the
fitting procedure presented in Sec. 4.

From Eq.~(\ref{A2}) the slope of the forward cone is

\begin{equation}\label{slope}
B(s,Q^2,t)=\frac{d}{dt}\ln|A|^2=2\left[b+\ln\left({s\over{s_0}}\right)\right]{\alpha'\over{1-\alpha_2
t}}+ 2b\,{\alpha'\over{1-\alpha_2 z}},
\end{equation}
which, in the forward limit, $t=0$ reduces to
\begin{equation}\label{slope1}
B(s,Q^2)=2\left[b+\ln\left({s\over{s_0}}\right)\right]\alpha'+
2b\,{\alpha'\over{1+\alpha_2 Q^2}}.
\end{equation}
Thus, the slope shows shrinkage in $s$ and antishrinkage in $Q^2.$

\section{Photoproduction- and DIS limits} \label{s3}

In the $Q^2\rightarrow 0$ limit the  Eq.~(\ref{A2}) becomes
\begin{equation}\label{amplitude4}
A(s,t)=-A_0e^{2b\alpha(t)}(-is/s_0)^{\alpha(t)}
\end{equation}
where we recognize a typical Regge-behaved photoproduction
(or, for $Q^2\rightarrow m_H^2,$ on-shell hadronic ($H$)) amplitude.
The related deep inelastic scattering structure function is
recovered by setting $Q_2^2=Q^2_1=Q^2$ and $t=0$, to get a typical 
elastic virtual forward Compton scattering amplitude:

\begin{equation}\label{amplitude5}
A(s,Q^2)=-A_0e^{b(\alpha(0)-\alpha_1\ln(1+\alpha_2
Q^2))}e^{(b+\ln(-is/s_0))\alpha(0)}\propto -(1+\alpha_2
Q^2)^{-\alpha_1}(-is/s_0)^{\alpha(0)}.
\end{equation}

 In the Bjorken limit, when both $s$ and
$Q^2$ are large and $t=0$ (with $x\approx Q^2/s$ valid for large
$s$), the structure function is given by:
\begin{equation}\label{F2}
F_2(s,Q^2)\approx {(1-x)Q^2\over{\pi\alpha_e}}\Im A(s,Q^2)/s,
\end{equation}
where $\alpha_e$ is the electromagnetic coupling constant and the
normalization is $\sigma_t(s)=\frac{4\pi}{s}\Im A(s,Q^2).$ The
resulting structure function has the correct (required by gauge
invariance) $Q^2\rightarrow 0$ limit and approximate scaling (in
$x$) behavior for large enough $s$ and $Q^2$.

It should be noted, however, that the Regge behavior has a limited
range of validity in $Q^2$. The smooth transition to DGLAP evolution was
studied in Ref.~\cite{CsJKLP}, while a relevant explicit model was
developed in Ref.~\cite{DJP}.

\section{Fits to the $ep\rightarrow e\gamma p$ data}\label{s4}

A standard procedure for the fit to the HERA data on DVCS~\cite{H1_05,ZEUS} 
based on Eq.~(\ref{A2}) has been adopted.  
A detailed analysis of the data would require a sum of
a Pomeron plus an $f$-Reggeon contribution:
\begin{equation}\label{amplitude6}
A=A^P+A^f.
\end{equation}

To avoid the introduction of too many parameters, given the
limited number of experimental data points, we use a single Reggeon term,
as already discussed in Sec. 2, which can be treated as an effective
Reggeon. The parameters
$\alpha(0)$, $\alpha_1$ and $\alpha'$ have been fixed
to 1.25, 1.0 and 0.38~GeV$^{-2}$ respectively and the values of the fitted 
parameters $A_0$ and $b$, described in Eq.~(\ref{A2}) are listed in 
Table~\ref{tab1}. The value of $\alpha'$ has been determined in an
exploratory fit with this parameter left free to vary 
between 0.2 and 0.4~GeV$^{-2}$.

The ZEUS measurements have been rescaled to the $W$ and $Q^2$ values of 
the H1 measurements.  
The mean value of $|t|$ has been fixed to 0.17 GeV$^2$ according with the H1 
measurements of the differential cross-section in  the range (0.1-0.8)GeV$^2$
for H1~\cite{H1_05} taking into account the value 6.02GeV$^{-2}$ for the slope 
$B$ as determined by the experiment.

\begin{table}[h]
\vspace{-.5cm}
\begin{center}
\begin{tabular}{|c||c|c|c|}
\hline
parameter &$\sigma_{DVCS}$ vs $Q^2$ & $\sigma_{DVCS}$ vs t & $\sigma_{DVCS}$ vs W\\
\hline
\hline
$|A_0|^2$ & 0.08 $\pm$ 0.01 & 0.11 $\pm$ 0.24 & 0.06 $\pm$ 0.01 \\
\hline
b & 0.93 $\pm$ 0.05 & 1.04 $\pm$ 0.91 & 1.08 $\pm$ 0.10 \\
\hline
$\chi^2/ndof$ & 0.57 & 0.15 & 1.15\\
\hline
\end{tabular}
\end{center}
\vspace{-.5cm}
\caption[]
{\small\it The values of the fitted parameters quoted in Eq.~(\ref{A2}).}
\label{tab1}
\end{table}

\begin{figure}[h]
\begin{center}
\includegraphics[clip,scale=0.60]{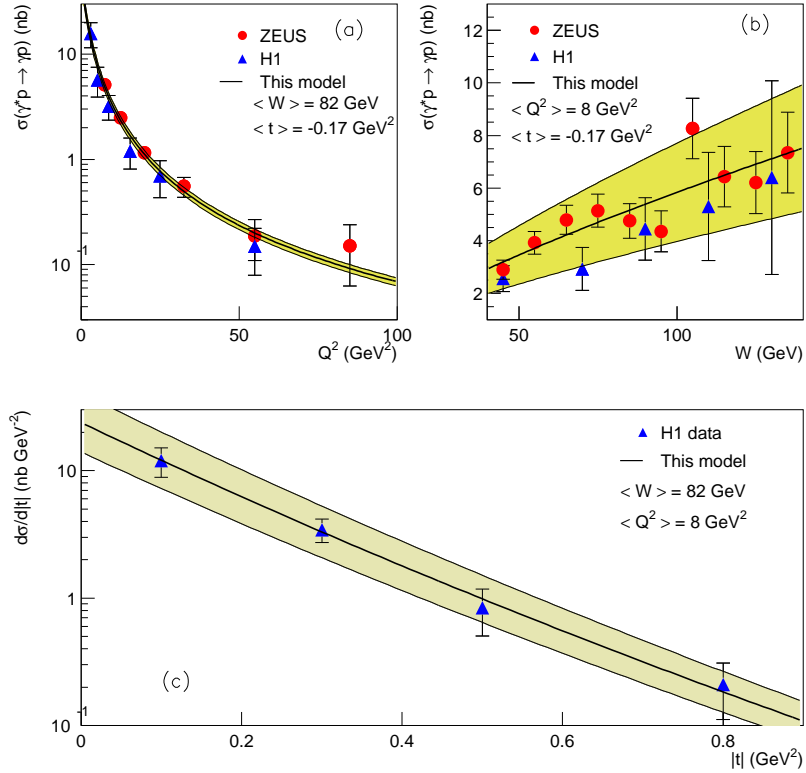}
\end{center}
\caption{\small\it{
The $\gamma *p\rightarrow \gamma p$ cross section
as a function of $Q^2$~(a), of 
$W$~(b) and the cross section differential in 
$t$~(c) measured by H1 and ZEUS experiments~\cite{H1_05,ZEUS}. The ZEUS 
measurements have been rescaled to the $W$ and $Q^2$ H1 values. The lines show
the results of the fits obtained from Eq.~(\ref{dsigma}) to the data.}}
\label{fig:fit}
\end{figure}

The results of the fits to the HERA data on
DVCS are shown in Fig.~\ref{fig:fit}. The cross-section
$\sigma (\gamma^*p\rightarrow \gamma p)$ as a function of $Q^2$ and 
$W=\sqrt s$ are presented respectively in Fig.~\ref{fig:fit}a and 
Fig.~\ref{fig:fit}b. The differential cross-section
$d\sigma (\gamma^*p\rightarrow \gamma p)/dt$, given by
\begin{equation}\label{dsigma}
{d\sigma\over{dt}}(s,t,Q^2)={\pi\over{s^2}}|A(s,t,Q^2)|^{2},\end{equation}
is presented in Fig.~\ref{fig:fit}c. 

The quality of the fits is satisfactory; in particular our model
fits rather well the cross-sections as a function of $Q^2$ and the 
cross-section differential in $t$.
Although the present HERA data on DVCS are well within the ``soft'' region, 
the model potentially is applicable for much higher values of $|t|$, 
dominate by hard scattering.

\vspace{-0.5cm}

\newpage
\begin{figure}[h]
\begin{center}
\hspace{-1.cm}
\includegraphics[clip,scale=0.7]{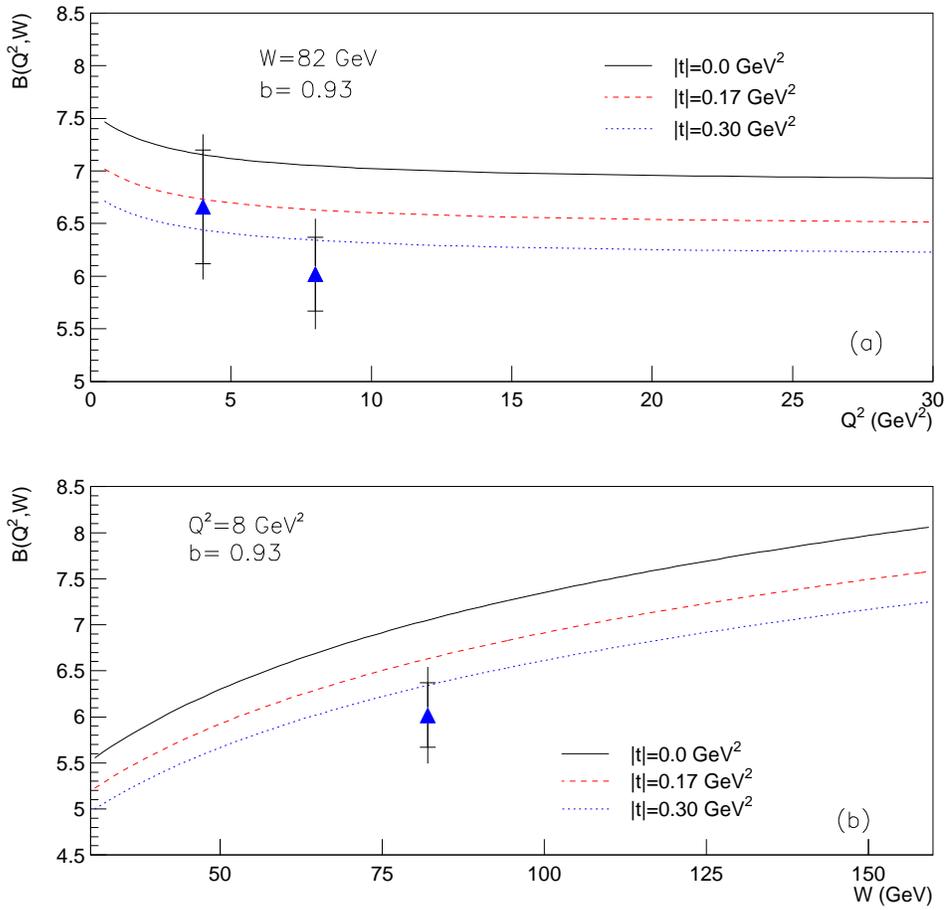}
\caption{\small \it {The $Q^2-$ and $s$ dependence of the local slope
described in Eq.~\ref{slope} (dotted and dashed line) and Eq.~\ref{slope1} 
(solid line). The triangles show the experimental measurements of H1}.}
\label{fig:fitb}
\end{center}
\end{figure}

Finally, Fig.~\ref{fig:fitb} shows antishrinkage in $Q^2$ and shrinkage in 
$W$ of the forward cone, according to Eqs.~\ref{slope} and~\ref{slope1}.
The curves are compared with the H1 experimental results.

\newpage
\section{Conclusions and discussion}
The model presented in this paper may have two-fold applications.
On one hand, it can be used by experimentalists as a guide. The
fits to the data could be improved, when more data are
available, by accounting for the Pomeron(s) and $f$-Reggeon
contributions separately as well as by using expressions for Regge 
trajectories which take exactly into account analyticity and 
unitarity. On the other 
hand, the model can be used to study various extreme regimes of
the scattering amplitude in all the three variables it depends on. For
that purpose, however, the transition from Regge behavior to QCD
evolution at large $Q^2$ should be accounted for. A formula
interpolating between the two regimes (Regge pole and asymptotic
QCD evolution) was proposed in Ref.~\cite{DJP} for $t=0$ only.
Its generalization to non zero $t$ value is possible by applying the ideas and
the model presented in this paper. The applicability of the model
in both soft and hard domains can be used to learn about the
transition between perturbative (QCD) and non-perturbative (Regge
poles) dynamics.

Independently of the pragmatic use of this model as a
instrument to guide experimentalists, given its explicit form, it
can be regarded also as an explicit realization of  the
corresponding principle~\cite{BK} of exclusive-inclusive
connection in various kinematical limits.

Last but not least, the simple and feasible model of DVCS
presented in this paper can be used to study general parton
distributions (GPD). As emphasized in Ref.~\cite{PF}, in the
first approximation, the imaginary part of the DVCS amplitude is
equal to a GPD. The presence of the Regge phase in our model can
be used for restoring the correct phase of the amplitude, for
which the interference experiments (with Bethe-Heitler radiation)
are designed.

{\bf Acknowledgments} L.L.~Jenkovszky is grateful to INFN and
Physics Departments of the Universities of Calabria and Padova for
their hospitality and support. S.~Fazio and R.~Fiore thank the
Dipartimento Galileo Galilei of the Padova University for the
hospitality extended to them during the conclusive stage of this
work.

\vfill \eject
\end{document}